\documentclass{article}
\usepackage{PRIMEarxiv}
\usepackage{pdfcomment}
\usepackage{tabularx}
\usepackage{amssymb}
\usepackage{soul}
\usepackage[utf8]{inputenc} % allow utf-8 input
\usepackage[T1]{fontenc}    % use 8-bit T1 fonts
\usepackage{hyperref}       % hyperlinks
\usepackage{url}            % simple URL typesetting
\usepackage{booktabs}       % professional-quality tables
\usepackage{amsfonts}       % blackboard math symbols
\usepackage{nicefrac}       % compact symbols for 1/2, etc.
\usepackage{microtype}      % microtypography
\usepackage{graphicx}
\usepackage{doi}
\usepackage{lipsum}
\usepackage{fancyhdr}       % header

\date{February 6, 2022}

\title{A Bibliometric Horizon Scanning Methodology for Identifying Emerging Topics in the Scientific Literature}

\author{
Artjay Javier* \\
Sciligent \\
Alexandria, Virginia, US \\
\texttt{artjay.javier@sciligent.com} \\
\And
Beth Masimore \\
Digital Science \\
Cambridge, Massachusetts, USA \\
\texttt{b.masimore@digital-science.com} \\
\And
John Chase\\
Digital Science\\
Cambridge, Massachusetts, USA\\
\texttt{j.chase@digital-science.com}\\
\And
F.G. Serpa\\
Sciligent\\
Alexandria, Virginia, USA\\
\texttt{gino.serpa@sciligent.com}\\
\And
John T. Rigsby\\
Naval Surface Warfare Center, Dahlgren\\
Dahlgren, Virginia, USA\\
\texttt{john.rigsby@navy.mil}\\
\And
Avory Bryant\\
Naval Surface Warfare Center, Dahlgren\\
Dahlgren, Virginia, USA\\
\texttt{avory.bryant@navy.mil}\\
\And
Jeffrey Solka\\
Naval Surface Warfare Center, Dahlgren\\
Dahlgren, Virginia, USA\\
\texttt{jeffrey.solka@navy.mil}\\
\And
Ryan J. Zelnio\\
Office of Naval Research\\
Arlington, Virginia, USA\\
\texttt{ryan.j.zelnio@navy.mil}\\
}

\begin{document}
\maketitle

%% ===========================================================================================
%%     ABSTRACT
%% ===========================================================================================

\begin{abstract}
A bibliometric methodology for scanning for emerging science and technology areas is described, where topics in the science, technology and innovation enterprise are discovered using \textit{Latent Dirichlet Allocation }, their growth rates are modeled using \textit{first-order rate kinetics}, and research specialization of various entities in these topics is measured using the \textit{location quotient}. Multiple interactive visualization interfaces that integrate these results together to assist human analysts are developed. This methodology is demonstrated by analyzing the last five years of publications, patents and grants ($\approx 14$ \textit{million documents}) showing, for example, that \textit{deep learning for machine vision} is the fastest growing area, and that China has a stronger focus than the U.S. in this area.

\end{abstract}

%% ===========================================================================================
%%     AUTHOR KEYWORDS
%% ===========================================================================================


%% ===========================================================================================
%%     START MANUSCRIPT
%% ===========================================================================================
\twocolumn
\section{Introduction}

% **********************************
%   FIGURE: Topic Map
% **********************************

\begin{figure*}[ht]
  \includegraphics[width=\textwidth]{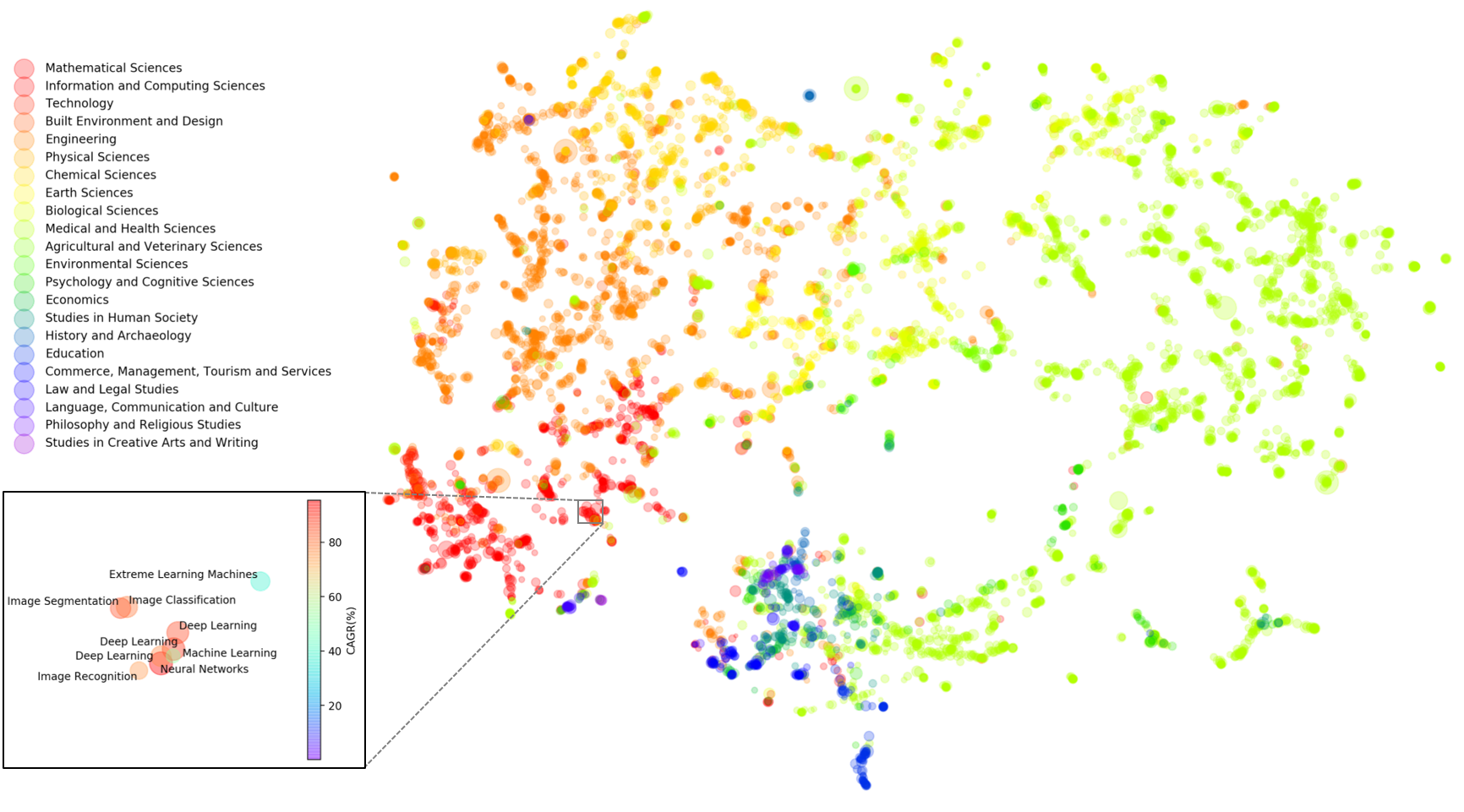}
  \caption{Map of 10,000 topics modeled with \textit{MALLET}\cite{mallet} using titles and abstracts from $\sim$14M publications, patents and grants from 2014-2018 drawn from the \textit{Dimensions} data set, with semi-transparent colors (with overlap leading to greater saturation) corresponding to \textit{Fields of Research (legend)}\cite{for}, marker size proportional to the number of documents, layout using \textit{UMAP}\cite{umap}, and final visualization with \textit{Python 3.5/matplotlib\cite{matplotlib}}. Inset: Magnified portion of the fastest growing region (colors based on Compound Annual Growth Rate (CAGR)).}
  \label{fig_topic_map_for}
\end{figure*}

Bibliometric horizon scanning\cite{hs_review} has long been used to assess and forecast trends\cite{bettencourt_emergence} in the scientific and technical literature\cite{chen_citespace} as well as measure the interaction between countries, organizations and people in vast interlinked collaboration\cite{coauthor_networks} and influence\cite{boyack_citation_analysis} networks. Horizon scans are often used by both public\cite{hs_eu} and private\cite{hs_industry} sector enterprises to strategically allocate their resources, whether that be investments, staff effort and hiring, or future policy decisions\cite{policy_roadmap}. The \textit{Organisation for Economic Co-operation and Development (OECD)} defines horizon scanning as \textit{a technique for detecting early signs of potentially important developments through a systematic examination of potential threats and opportunities, with emphasis on new technology and its effects.}\cite{hsdef_oecd} Our methodology addresses these challenges by (1) systematically ingesting a broad collection of millions of recently published science and innovation documents, (2) examining rapidly growing research and innovation areas, emphasizing small research areas that may receive less attention, (3) identifying key actors who may present both threats and opportunities, and (4) integrating human interpretation to help assess the impact of these new technologies. A key motivation informing our approach is the need for transparency and clarity to facilitate the human-machine interaction by assisting the analyst/subject matter experts examining the results of the bibliometric horizon scan to synthesize a cohesive story and assemble the evidence to support it. Therefore data visualization, simple intuitive metrics, and rapid screening and filtering techniques are central themes in our methodology.

\subsection{Analyzing Technical Concepts}

\begin{figure*}[ht]
\includegraphics[width=\textwidth]{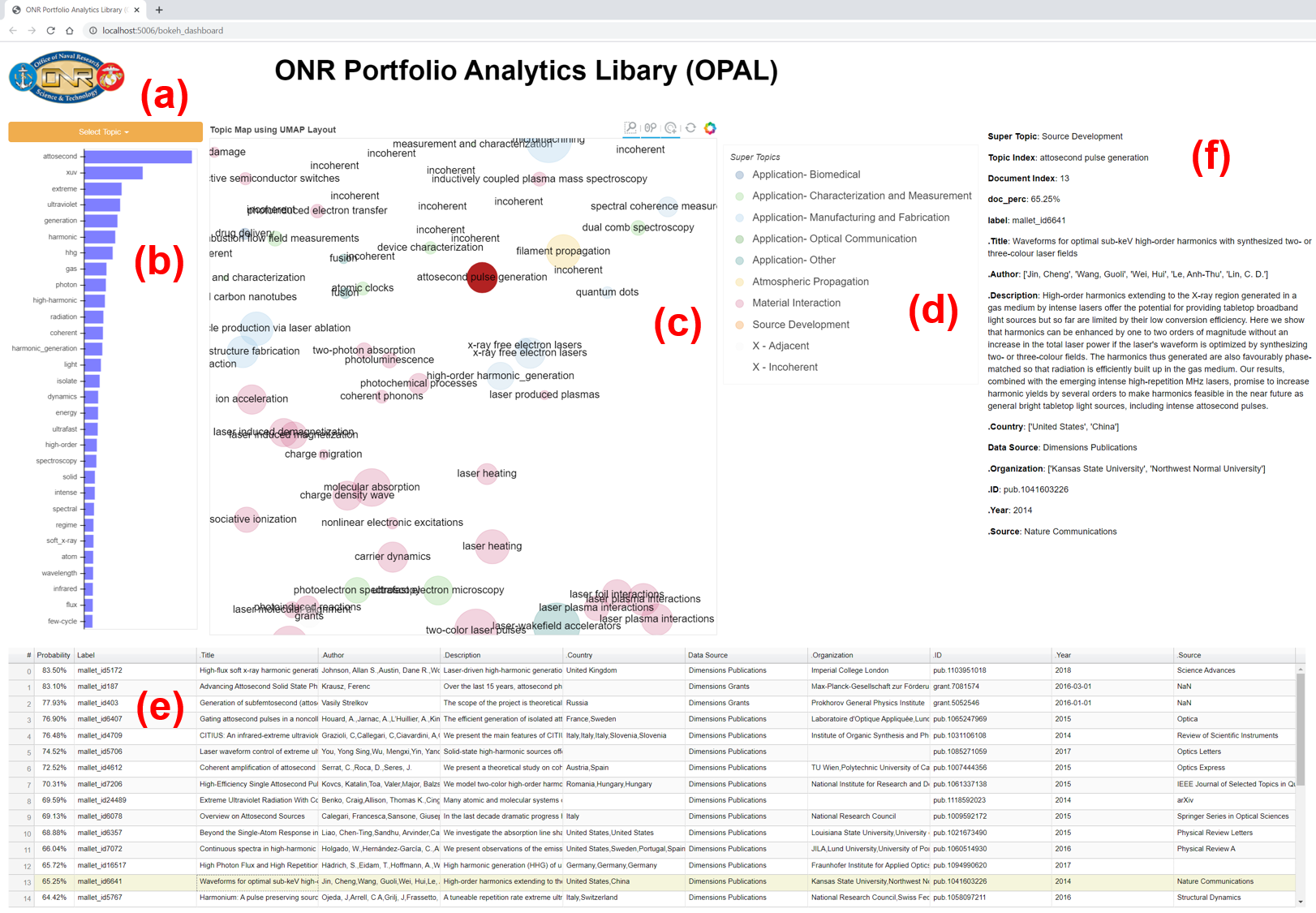}
\caption{A screenshot of the interactive \textit{Bokeh} dashboard used by analysts to label and explore the content of the corpus as part of the \textit{ONR Portfolio Analysis Library (OPAL) (see footnote in corresponding text for more information)} with the following elements: (a) a drop-down menu listing topics by index, (b) terms in descending order of relative frequency for the selected topic, (c) an interactive, scalable view of the topic map using the UMAP layout, and shown with the analyst-assigned labels, (d) legend for the topic map showing super topics aggregated by different colors, (e) a document browser showing documents with a high topic probability in the selected document, (f) a detail widget showing features of both the topic and document selected.}
\label{fig_dashboard}
\end{figure*}

The challenge of efficiently and thoroughly analyzing millions of scientific documents can be addressed with modern machine learning methods, specifically unsupervised learning which exploits patterns inherent in a collection of documents, and requires no \textit{a priori} knowledge, no pre-existing taxonomic structure, and no human-labeling of documents. While traditional hard clustering methods (\textit{i.e. k-means clustering\cite{cluto}}) group documents into clusters that limit documents to possessing membership in only one cluster, soft clustering methods \textit{i.e. Latent Dirichlet Allocation (LDA)}\cite{blei_lda} allow documents to be members of multiple topics, and likewise for topics to be composed of multiple documents. This additional flexibility is a more accurate reflection of real documents, and affords documents with strongly mixed themes an additional degree of freedom during optimization. However, a drawback of unsupervised learning is that the topics are not immediately human readable, consisting typically of lists of words with various probabilities. While recent methods\cite{glanzel_labeling} have made this topic labeling process more automatic, more consistent and nuanced labels can be potentially created by experienced analysts if they are given the tools to thoroughly and efficiently evaluate the topic model. While multiple open source tools provide separate avenues to visualize topic models\cite{termite,hierarchie,pyldavis}, none comprehensively give the analyst (1) a 2D map of these topics revealing clusters of related topics (like the one shown in Figure \ref{fig_topic_map_for}), (2) a coherent view of both the \textit{term/topic} and \textit{document/topic} distributions, (3) a way to apply labels and aggregate topics into custom categories to further shape the appearance of the topic map, (4) reports of objective measures of the topic quality of each topic, and (5) a single web-based browser interface that performs all of the above. We discuss results from the use of one such interface (shown in Figure \ref{fig_dashboard}) developed for this purpose and whose agile development benefited from continuous feedback from analysts.

% **********************************
%   FIGURE: Bokeh Dashboard Screenshot
% **********************************
% **********************************

\subsection{Detecting Emerging Science and Technology}

Methods to detect emerging sciences and technologies must be sensitive to "weak signals\cite{weak_signals}"  areas too early in their maturity to be strong enough to compete with other more mature topics, but growing at such a fast and consistent rate that their future position of dominance is very likely. It has been commonly observed that counting the number of scientific publications and patents published per year reveals a rising, superlinear trend. However, detailed modeling of the entire scientific enterprise is a complex undertaking and beyond the scope of this investigation. To simplify the problem, we model the observed superlinear growth with the simplest possible formulation by using first-order rate kinetics: the rate of growth/decay$\left(\frac{dN}{dt}\right)$ is proportional to the number of publications ($N$), shown in Equation \ref{eq_rate}. 

\begin{equation}
    \frac{dN}{dt}=kN
\label{eq_rate}
\end{equation}

When integrated, this model predicts an exponential growth or decay in publications, over time, as shown in Equation \ref{eq_rate_int}, where $N=N_{0}$ when $t_{0}=0$.

\begin{equation}
    N=N_0e^{kt}
\label{eq_rate_int}
\end{equation}

To improve the human interpretability of the first-order rate equation, we rearrange it to represent the familiar value of the \textit{Compound Annual Growth Rate (CAGR)} popularized in finance, and sometimes called the \textit{Year-Over-Year (YOY)} percentage increase relative to the prior year. CAGR can be directly derived
    \footnote{Consider a full growth interval ($\Delta t = t_f-t_i=1$) where $N_i$ has grown by a percentage amount, CAGR, to $N_f=(1+CAGR/100)N_i$, then insert into Equation \ref{eq_rate_int}, then rearrange for CAGR to derive Equation \ref{eq_cagr}.}
from the \textit{rate constant} ($k$) as:

\begin{equation}
    CAGR(\%)=100 \left( e^{k}-1 \right)
\label{eq_cagr}
\end{equation}

CAGR is a proxy for the rate of emergence, emphasizing not the overall strength of any signal, but how fast that signal grows over time. This lack of dependency on the overall signal has the potential to make it more sensitive to so-called "weak signals." By focusing, for example, on topics that have high CAGR but low document frequencies, finding emerging topic areas still in their infancy is possible. More details on the statistical testing of this model appear in the Supplemental Information.

\subsection{Measuring Area Specialization}

For understanding opportunities and threats, the actions of various entities in different arenas can often produce a coherent picture that describes their strengths and weaknesses. However, raw scores of their activities in different areas can be complicated by (1) scale, wherein some entities have more influence and resources than others, thus allowing smaller players to go unnoticed, and (2) the tendency for many entities to favor the same arenas resulting in their patterns to look so similar that their potentially important subtle differences are lost.

To address these issues, the \textit{Location Quotient\cite{lq} (LQ)} is employed as the prime evaluation metric for determining an entity's focus. 
Denoting $n_{i,j}$ as the activity of an entity $j$ in an area $i$, we express the focus of entity $j$ on area $i$ as $LQ_{i,j}$  (Equation \ref{eq_lq}) where the numerator is the activity of an entity normalized by the sum of its areas of interest ($\frac {n_{i,j}} {\sum_{k}{n_{k,j}}}$) and the denominator evaluates the same ratio but for all entities ($\frac {\sum_{l}{n_{i,l}}} {\sum_{k,l}{n_{k,l}}}$).

\begin{equation}
    LQ_{i,j} = \frac {\frac {n_{i,j}} {\sum_{k}{n_{k,j}}}}
                     {\frac {\sum_{l}{n_{i,l}}} {\sum_{k,l}{n_{k,l}}}  }
\label{eq_lq}
\end{equation}

LQ addresses the above concerns in the following ways: First, because LQ is normalized against the sum of an entity's proportions in all categories, it enables scale-free comparisons between entities. Second, because it is normalized against the relative proportions of all entities over all categories, it receives a boost in its sensitivity against this global baseline. Finally, the LQ is easily interpretable: For a category where $LQ < 1$, the entity is underrepresented relative to all others, whereas $LQ=1$ categories are where the entity has the same proportions to its peer average, and when $LQ >1$ the entity is more represented in these categories than the global average. However, the boosted sensitivity of the LQ has its drawbacks. For example, since it is doubly normalized, it can cause high error rates at low counts. As a mitigation, these errors are assessed using typical error propagation formulas\cite{error_propagation}, which aids in screening potentially spurious LQ values.

\section{Materials and Methods}

\subsection{Source Database \& Vocabulary Preparation}

The \textit{Dimensions}\cite{dimensions} data sets were used exclusively in this work. Scientific publications with a publication date, or patents with a granted date, or grants with a start date, between 2014 to 2018 were used in this analysis. This resulted in $11,517,625$ publications, $937,369$ grants and $1,436,876$ patents. Titles and abstracts were extracted from the documents and processed using the following steps, which included the use of a vocabulary preparation methodology\cite{mimno_phrasing} based on topic-based phrases. Complete details appear in the Supplementary Information.

\subsection{Topic Modeling}

The LDA implementation in \textit{Machine Learning for Language Toolkit (MALLET)}\cite{mallet} was selected because of its use of Gibbs sampling for robust optimization, multi-threading for fast computation, topic-based phrase generation to assess topic content, and automatic generation of topic diagnostic output for objective assessment of topic model quality. Based on prior experience with this corpus and other similar corpora, it was decided to fit the corpus to $10,000$ topics to obtain distinct topics which minimize mixing between unrelated topics. 

The \textit{topic/document} matrix is a side-product of the topic-modeling process that reports a probabilistic topic value $(t_{i,j})$ for every topic$(i)$ to every document$(d_{j})$ for all documents$(j)$. Since this value is normalized for any given document ($\sum_{i}{t_{i,j}}=1$), we carried out all subsequent calculations over topics that require a count of documents as a sum of these topic probabilities ($\sum_{j}{t_{i,j}}$)
     \footnote{Furthermore, when this is summed over all documents$(j)$ and all topics, this produces: $\sum_{i,j}{t_{i,j}}=D$, where $D$ is the sum of all documents in the corpus. This is both theoretically true and empirically observed in this work}. 
For simplicity, we report sums obtained in this way as \textit{Documents} or \textit{document counts}, but as described above, they are not necessarily whole numbers
     \footnote{For any given topic, \textit{fractional document counting} will typically undercount \textit{whole document counting} by a factor roughly proportional to the average number of words in the text. In this case, since we use the titles and abstracts, together they have a sum of total words of order $\sim 10^{2}$. The cause of this undercounting is that documents contain many words that are not specific to a particular topic and since topic frequencies are based on the word count in a document, they will be proportionately affected.}.

\subsection{Metrics}

\subsubsection{Compound Annual Growth Rate (CAGR)}

CAGR was calculated by summing the number of \textit{Documents} (as described in the previous section) per topic per year to create a time series. Nonlinear least-squares curve-fitting was then carried out using the \textit{LMFIT}\cite{lmfit} Python package to obtain optimized fit parameters and their corresponding errors for all $10,000$ topics. This technical improvement addresses issues inherent in the conventional CAGR formula which uses only two data points
    \footnote{The traditional CAGR formula relies only on the initial point $(t_{0},N_{0})$ and the final point $(t_{n},N_{n})$ such that: $CAGR(t_{0},t_{n})=100 \left( \left( \frac{N_{N}}{N_{0}}\right)^\frac{1}{t_{n}-t_{0}}-1\right)$},
and inherently provides no estimate of error. By fitting the model to all the data points more robust and accurate calculation are achieved. Furthermore the fit parameter errors provide insight on how closely, or loosely, the data can be described by the model.

% **********************************
% FIGURE:  CAGR CALIBRATION
% **********************************
\begin{figure}
\includegraphics[width=\columnwidth]{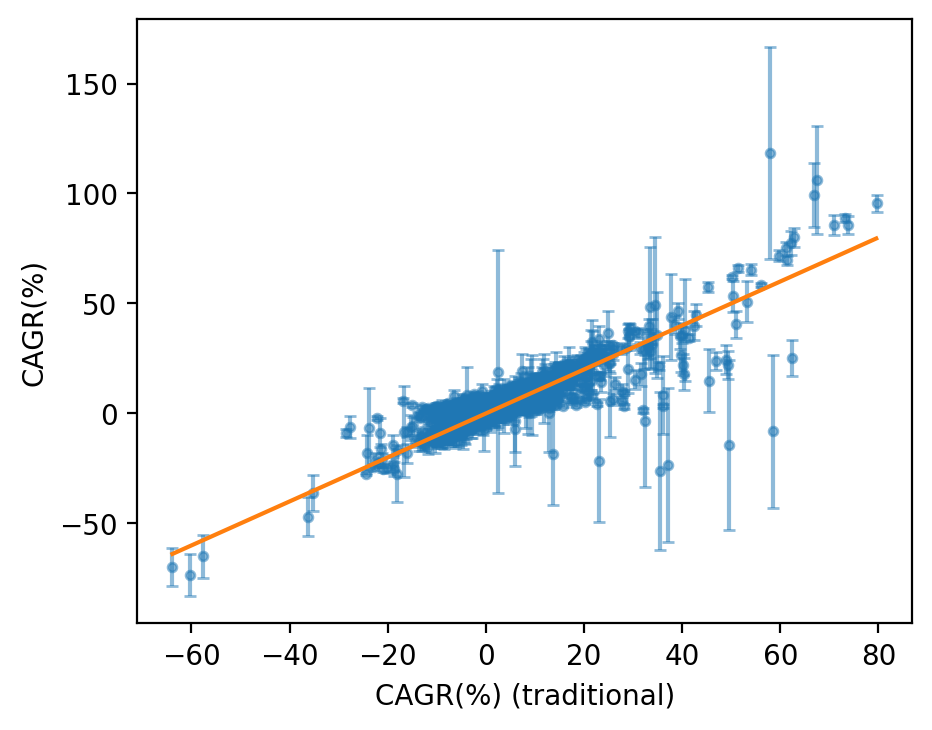}
\caption{A calibration-style plot showing \textit{CAGR(\%)} calculated using nonlinear regression with Equation \ref{eq_cagr} on the \textit{y-axis} and plotted with fit errors as error bars, and the traditional two-point CAGR formula on the \textit{x-axis}, \textit{CAGR(\%) (traditional)}. Calibration line (orange): y = x}
\label{fig_cagr_calibration}
\end{figure}
% **********************************

The value of using nonlinear regression versus the traditional CAGR formula\cite{cagr} can be seen in Figure \ref{fig_cagr_calibration}. Several points with large errors appear outside of the calibration line - points that could not be distinguished from the others without an error estimation. Below the calibration line are points that appear with a larger apparent CAGR based on the two-point formula. These points would be regarded as false positives if a CAGR screen was used with the two-point formula. Above the calibration line are topics that would have received a lower ranking based on the two-point formula, or points that would have been missed in a CAGR screen. At high CAGR (right hand side of plot) there is a noticeable positive (upward) deviation from the calibration line, indicating a general underestimation of CAGR by the traditional formula.

\subsubsection{Location Quotient (LQ)}

LQ was calculated with entities including the country, sponsor, organization, individual, or data source. Each entity's representation in a topic was determined using \textit{document counts} with the LQ formula in Equation \ref{eq_lq}.

\subsection{Visualization \& Analyst Labeling}

To visualize the results of the topic map, several \textit{multi-dimensional scaling (MDS)} algorithms were evaluated including \textit{Principal Component Analysis (PCA)} and \textit{t-Distributed Stochastic Neighbor Embedding (t-SNE)}. \textit{Uniform Manifold Approximation and Projection (UMAP)}\cite{umap} provided the best (and fastest) visualization based on the criteria that topics closely resembling each other would be located near each other as topic clusters. This clustering behavior can be controlled in UMAP to favor local structure over global structure by constraining the local neighborhood prior to optimization
    \footnote{The parameter \textit{n\_neighbors} controls this local versus global tradeoff in \textit{UMAP} and was set to 15 (the default value) for this study.}.
Custom interactive dashboard interfaces using \textit{Plotly Dash}\cite{dash} or \textit{Bokeh}\cite{bokeh} were developed for visualizing \textit{MALLET's} outputs
    \footnote{These include the optimized Gibbs state file, the topic diagnostics XML file, and the topic-phrase report XML file.}. 
These visualizations were strongly inspired by \textit{pyLDAvis}\cite{pyldavis}. To facilitate the analyst labeling process, interactive dashboards were employed using \textit{OPAL}
    \footnote{\textit{ONR Portfolio Analysis Library (OPAL)} is a Python-based Government Off the Shelf (GOTS) product developed by \textit{Sciligent, LLC} and is made available only to U.S. Government Employees and their direct support contractors.}. 
A screenshot of an example dashboard appears in Figure \ref{fig_dashboard}. The topic map is zoomable and displays topic information upon mouseover. It also labels each topic with the analyst-selected name, and colors the topic according to a \textit{supertopic} scheme devised by the analyst (described below). Upon clicking on any topic, the term distribution is shown as a horizontal bar chart, which is useful for determining the relative dominance of various words in the topic: for example, topics with a narrow distributions behave more like single terms. While many modern topic model visualizations\cite{pyldavis,hierarchie,termite} leave out the topic/document matrix, we implemented visualizing this as a document browser, sorted in descending order by the fractional composition of that topic within each document. This is useful for determining if that topic is strongly represented in a single document or broadly distributed at very low fractional levels across multiple documents. Selecting any document in the browser then displays document information for further inspection, including the full abstract/description which can be useful for the analyst to understand, interpret or give context as to why the topic is represented in those documents. Last, the proximity of that topic to other topics in the topic map give some indication of the broader category of that topic. These separate elements contribute to an exploratory workflow that the analyst uses to quickly assign a topic label, to understand the structure and content of the corpus, and to discover patterns within the corpus that may not be apparent from document sampling or viewing existing categories in isolation. For example, interdisciplinary areas and applied research areas are often found in close proximity to core research areas in the topic map visualization.

Using these dashboard interfaces, analysts labeled each topic index with two names: \textit{Topic Name} and \textit{Super Topic Name}. The \textit{Topic Name} is a human-readable label, typically 2-3 words, that attempts to concisely capture what the topic is about. The \textit{Super Topic Name} is assigned based on the goal of aggregating multiple related topics into a single umbrella term, with the goal of limiting the total number of \textit{Super Topic} categories to a manageable quantity $(< 20)$. Not all topics contain technology-specific information relevant to this study and were ignored
    \footnote{These omitted topics include those that are non-technical \textit{(i.e. containing administrative or marketing language)}, are mixtures of two or more unrelated topics, or are trivial \textit{(i.e. the names of all amino acids, or a list of colors)}, or are not indicative of a specific technology area \textit{(i.e. clinical trials)}. Examples appear in the Supplementary Information.}.
For static output and semi-automated batch processing of clusters of topics, \textit{matplotlib/Python} was used with \textit{adjustText}\cite{adjust_text} to improve readability, such as the one shows in the inset of Figure \ref{fig_topic_map_for}.

% =================================================
%           RESULTS
% =================================================

\section{Results}

\subsection{Topic Diagnostics Output}

Topic Diagnostics were examined from the \textit{MALLET} output to assess the quality of both the overall model as well as individual topics. The \textit{coherence} is a useful measure of how well the underlying documents are represented by the highest frequency terms of that topic and a potentially good indicator of topic quality\cite{coherence}. From prior experience, a coherence of $\gtrapprox -1000$ is typically indicative of well-formed topics. The vast majority of topics $(>92\%)$ are above this threshold \textit{(See Supplemental Information)}.

\subsection{Topic Map Topology}

\subsubsection{Global Structure}

While the 2D Cartesian axes need not necessarily follow any particular pattern, a speculative observation consistent with repeated permutations presents some useful heuristics: One axis (in this permutation, the vertical axis) can be interpreted as a spectrum of \textit{scale} starting at the top with microscopic particles and ending at the bottom with societies of people. The other axis (in this permutation, horizontal axis) seems to correspond to \textit{complexity}, starting at the left with abstract concepts and lifeless particles and ending at complex organisms on the right. Importantly, this is not a scheme that was imposed, but which arises naturally from multidimensional scaling. It is also \textit{not} unique to UMAP, or the \textit{Dimensions} data set, or even our software development cycle
    \footnote{We have observed it using \textit{both} in PCA and t-SNE, in other S\&T literature databases such as the \textit{Web of Science}, and simply using \textit{pyLDAvis} for visualization with any of the above.}.

\subsubsection{Local Structure}

The local morphology exhibits long, distinct, strand-like filaments. Topics align themselves in these one-dimensional structures with significant white space separating them. This strong clustering pattern can be emphasized with \textit{UMAP}, as opposed to other multi-dimensional scaling algorithms such as \textit{PCA}, which allows clusters of topics to be easily distinguished. This control over the local clustering also allows for the visualization of areas which are closely related, such as the relationship between \textit{machine vision} and \textit{deep learning} as shown in Figure \ref{fig_topic_map_for} inset.

\subsubsection{Source Localization}

% **********************************
% FIGURE:  Pubs, Patents, Grants
% **********************************
\begin{figure*}
\includegraphics[width=\textwidth]{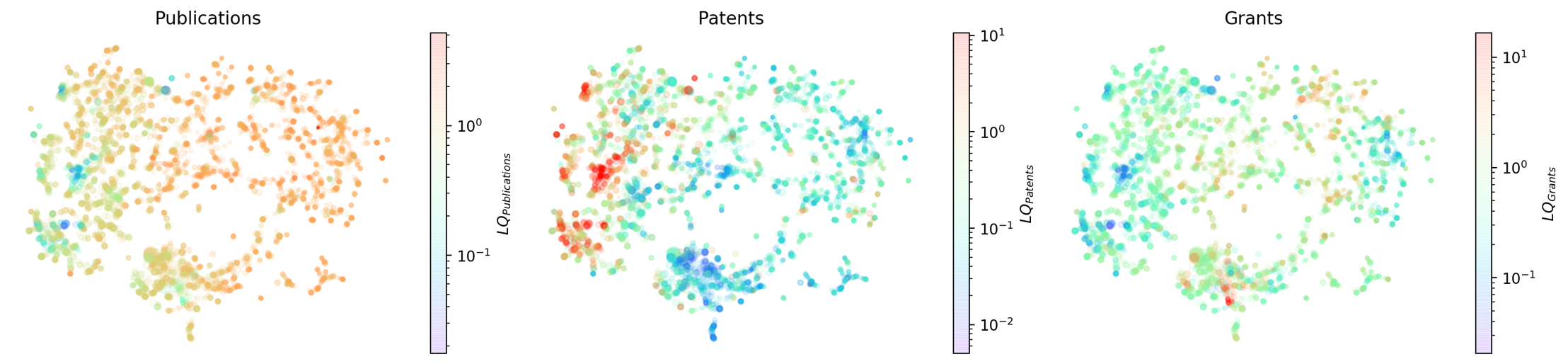}
\caption{Topic Map of $10,000$ topics with the same layout parameters as Figure \ref{fig_topic_map_for}. Color is proportional to the \textit{Location Quotient (LQ)} of the data source shown in the colorbar to the right of each source: \textit{Publications (left), Patents (middle), or Grants (right).}}
\label{fig_pubs_pats_grants}
\end{figure*}
% **********************************

A key concern of merging together multiple sources is that the inherent tendencies toward corpus-specific vocabularies may cause an attempt to combine them to result in undesirable phase segregation
    \footnote{Some of these effects were mitigated by the removal of problematic terms. For example, adding "claim," a term frequently but non-specifically in the patent literature, to the stopword list}. 
Our approach allows us to visualize the magnitude of any such segregation from the topic map topography: Figure \ref{fig_pubs_pats_grants} recolors Figure \ref{fig_topic_map_for} according to the LQ of the source data: publications, patents or grants. As can be seen, \textit{publications} are distributed evenly with slightly stronger focus in \textit{biomedicine and biology}. Not surprisingly, \textit{patents} are strongly focused in \textit{engineering and technology}. \textit{Grants} are broadly distributed but have a specific focus in the \textit{humanities}
    \footnote{Not necessarily due to increased funding, but likely to arise from less patent and publication activity. Grants are also not limited to traditional academic areas but also fund, for example, museums and other learning centers.}.
It can be seen that there is no strong segregation \textit{(i.e. separated clusters with large intervening white space)}, indicating the shared technical terminology between these corpora is stronger than differences in the language usage between them, thus resulting in most topics possessing mixtures of different sources.

\subsection{Labeling \& Aggregating the Fastest Growing Topics}

% ================
%    TABLE: Fastest growing CAGR topics 
% ================
\begin{table*}
\small
\caption{The top 10 fastest growing (highest CAGR) topics, including the analyst-assigned labels (\textit{Topic Name, Super Topic Name}), the \textit{Topic Size} (by virtual document count), the \textit{Compound Annual Growth Rate (CAGR)}, the topic \textit{Coherence}, and the top 5 highest frequency terms in that topic.}
\label{table_topics}

\begin{tabular}{rlllcc}
\toprule
 \textbf{Topic} &    \textbf{Topic Name} & \textbf{Super Topic} &                                         \textbf{Top Terms} &  \textbf{Coherence} & \textbf{CAGR(\%)} \\
 \textbf{Index} &                        & \textbf{Name}                                     \\

\midrule
                 4102 &             Blockchain &                Technology &   blockchain, transaction, bitcoin, currency,  &                -439 &            $106 \pm 25$ \\
                  676 &        Neural Networks &           Neural Networks &           neural\_network, train, cnn, deep\_learning, &                -365 &              $95 \pm 4$ \\
                  350 &          Deep Learning &           Neural Networks &            deep\_learning, deep, train, neural\_network, &                -412 &              $89 \pm 2$ \\
                 4403 &          Deep Learning &           Neural Networks &  deep, neural\_network, train, deep\_learning,  &                -428 &              $86 \pm 4$ \\
                 9356 &     Image Segmentation &            Machine Vision &             image, train, cnn, segmentation, &                -356 &              $86 \pm 5$ \\
                 9392 &   Image Classification &            Machine Vision &                     image, train, cnn, deep, convolutional &                -342 &              $80 \pm 4$ \\
                 9759 &          Deep Learning &           Neural Networks &                   train, deep\_learning, image, deep, learn &                -402 &              $77 \pm 6$ \\
                 9252 &      Image Recognition &            Machine Vision &             image, cnn, recognition, classification, train &                -354 &              $75 \pm 3$ \\
                 3096 &       Machine Learning &          Machine Learning &              train, learning, learn, algorithm, generative &                -420 &              $53 \pm 7$ \\
                 1053 &  Perovskite Solar Cell &               Solar Cells &                  perovskite, solar\_cell, pbi, halide, film &                -349 &              $51 \pm 9$ \\
\bottomrule
\end{tabular}

\end{table*}
% ================

Since CAGR is a useful screen for allocating analyst labor toward potentially emerging topical areas, analysts labeled the top 200 fastest growing (highest CAGR) topics. The top 10 identified in this way shown in Table \ref{table_topics}. Then analysts labeled each topic with a \textit{Topic Name} consisting of 1-4 words and a \textit{Super Topic Name} drawn from a controlled list of categories, such that each Super Topic is a broader category for which other topics would be held under, as shown in Table \ref{table_supertopics}. This process created human-readable labels for the topics and the opportunity for analysts to apply a custom taxonomy. Here, we elected to group topics without any such agenda, and pursued only the goal of reducing the number of aggregated topics to less than an arbitrarily selected 20 super topics.

\subsection{Rapidly Emerging Science \& Technology}

% **********************************
% FIGURE:  CAGR
% **********************************
\begin{figure}
\includegraphics[width=\columnwidth]{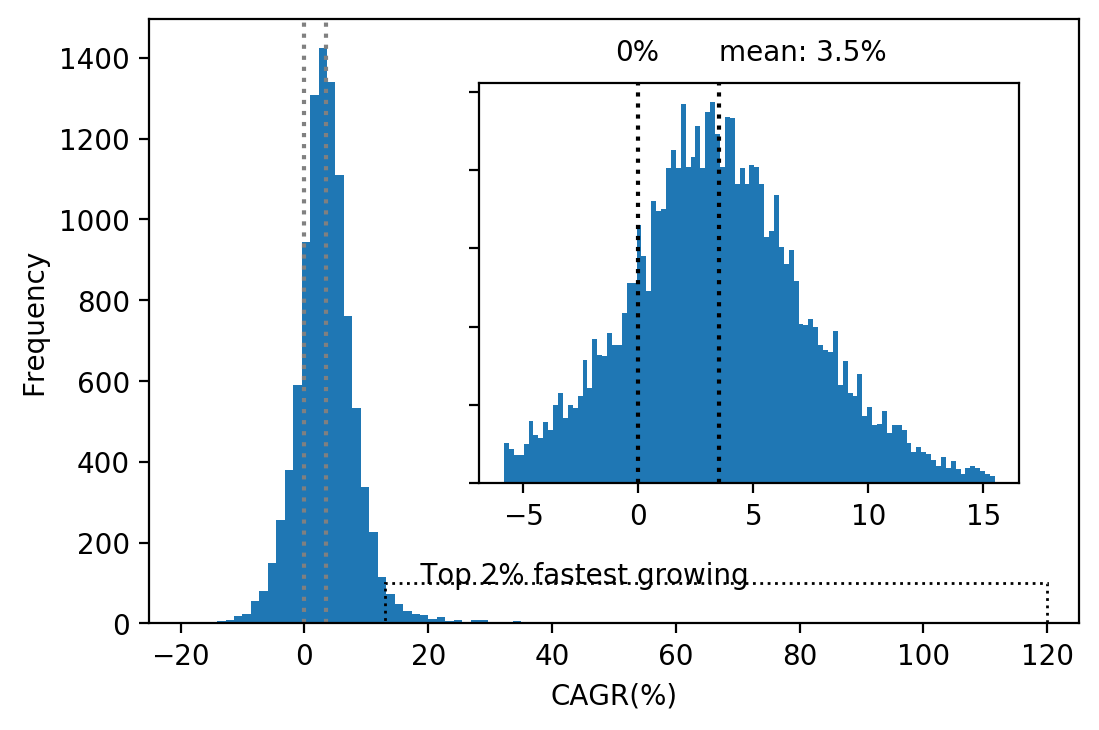}
\caption{A histogram of \textit{Compound Annual Growth Rate (CAGR)} for all $10,000$ topics modeled. Reference lines are shown for the mean growth rate and 0\% growth rate, as visual indicators revealing inflationary growth by the majority of topics. Box drawn shows the top 2\% of topics that are analyzed in further detail in this work. \textit{INSET}: Expanded view showing shape of distribution.}
\label{fig_cagr}
\end{figure}
% **********************************

The CAGR values for all 10,000 topics are plotted as a histogram shown in Figure \ref{fig_cagr}. The CAGR plot has the shape of a normal distribution with the mean$\pm$standard deviation of $3.5\%\pm3.9\%$ , removing outliers by excluding points more than three standard deviations away. The mean of the standard errors of CAGR (determined from the nonlinear regression) is $1.4\%$. Since the the mean standard error is less than the standard deviation of CAGR, we conclude that the spread in the distribution is statistically significant since the width of the CAGR peak is more than double that of the mean standard errors
    \footnote{Since the standard deviation of CAGR $(\sigma=3.9)$ was obtained from a large sample size (10,000) the chances that the true standard deviation of the distribution is the same as the mean error $(\epsilon=1.4)$ is negligible based on the results of statistical testing \textit{(p-value=0)}.}. 
However, because the sizes of these errors is still somewhat comparable, the shape of the peak is not an exact representation of the underlying CAGR distribution, but likely to be convoluted with the Gaussian error function leading to line broadening and a potential loss of distinguishing features.

% **********************************
% FIGURE: Counts vs CAGR
% **********************************
\begin{figure}
\includegraphics[width=\columnwidth]{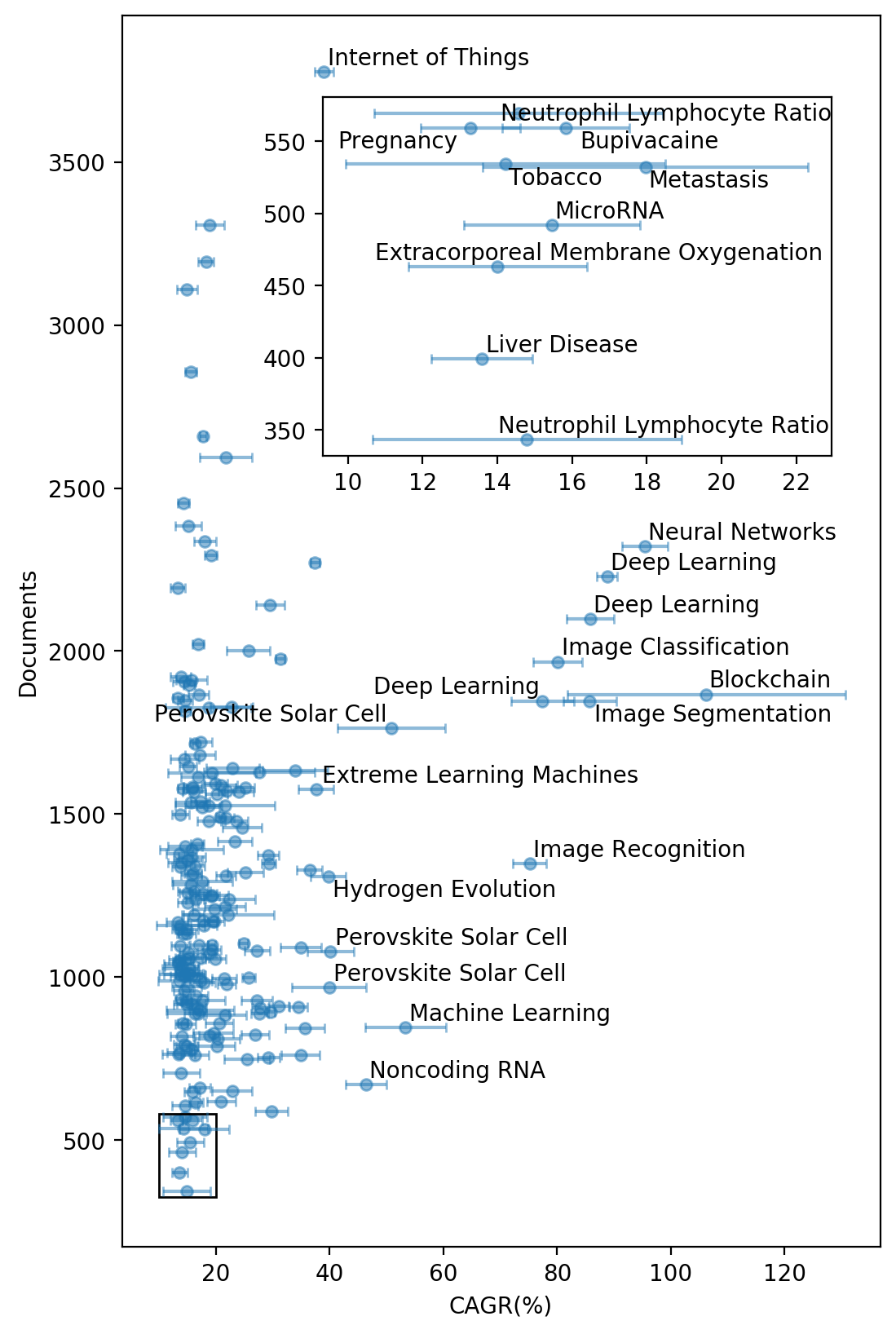}
\caption{For the top 200 ($2\%$) fastest growing topics, the topic size (based on the number of documents) and the \textit{Compound Annual Growth Rate (CAGR)}, with associated fit errors shown as error bars, are plotted. The 15 fastest growing topics are labeled with the analyst-selected \textit{Topic Name}. \textit{INSET}: Expanded plot of the box drawn at the bottom left-hand side of the main figure.}
\label{fig_counts_cites_cagr}
\end{figure}
% **********************************

\subsection{Evaluating Weak Signals}

Another useful factor for analytic consideration is the size of the topic, which provides an additional opportunity to screen for topics that are smaller in size, and potentially less well-known. These low-frequency count topics are the "weak signals" that are often ignored in traditional analyses that track the dominant areas and dominant entities. However, simply sifting through these low frequency count topics alone can be a tedious and time-inefficient task, as the large majority are unlikely to be emerging topics areas. However, combining this with the CAGR screen helps highlight which of these areas should be focused on: their rapid growth is a leading indicator that their limited awareness and small footprint today may be drastically different tomorrow. In Figure \ref{fig_counts_cites_cagr} we plot the top 200 topics based on their size and growth rate. This plot demonstrates how low document count, fast growing topics can be identified. Here, we can see multiple topics in \textit{deep learning} of significant size growing very rapidly. Of lower size and growth rate are \textit{perovskite solar cells}, still significant but not in the same class as the deep learning topics. Finally, if the lowest signals are examined (which are still growing significantly within the top $2\%$ of fastest growing topics), \textit{neutrophil to lymphocyte ratio} and other health-related topics can be interpreted as the small, rapidly emerging topics. Some topics such as \textit{bupivacaine} are not well known, and may not necessarily be an emerging research area and ultimately require analyst assessment to determine the cause of their growth. Thus, this approach does not decisively identify emerging topics, but instead helps narrow the topics requiring technical due diligence, which more efficiently directs analyst effort.  Another important limitation arises from analyzing only the top $2\%$ of fastest growing topics is that closely related, but slower growing portions are not included when aggregating the topic size. For example, the super topic group \textit{Neural Networks} contains only the fastest growing areas, but many other slower growing areas can be placed in this grouping.

\subsection{Major Players \& Rising Competitors}

% ==================
%  TABLE
% ==================
\begin{table*}
\small
\caption{The top 200 (2\%) fastest growing topics aggregated up to the top 20 \textit{Super Topics}. \textit{Topics} are listed in descending order of frequency, with CAGR and associated error also listed.}
\label{table_supertopics}

\begin{tabular}{cll}
\toprule
\textbf{CAGR(\%)} & \textbf{Super Topic Name} &                                                                   \textbf{Top Topics} \\
\midrule
             $77 \pm 3$ &           Neural Networks &                             Deep Learning, Extreme Learning Machines, Neural Networks \\

             $60 \pm 4$ &            Machine Vision &  Image Classification, Hyperspectral Image Classification, Image Recognition,  \\
             $37 \pm 5$ &               Solar Cells &                                                                 Perovskite Solar Cell \\
             $23 \pm 1$ &                 Catalysts &  Photocatalysts, Catalysts, Oxygen Reduction, Photocatalytic Water Splitting,  \\
             $22 \pm 1$ &                Technology &  Mobile Computing, Text Analytics, Cloud Computing, Internet of Things,  \\
             $21 \pm 1$ &                    Cancer &  Neutrophil Lymphocyte Ratio, Cancer Immunotherapy,  \\
             $21 \pm 0$ &           Gene Expression &  MicroRNA, Long Noncoding RNA, CRISPR/Cas9, Exosome, Cell Free DNA,  \\
             $21 \pm 2$ &             Manufacturing &  Additive Manufacturing, Manufacturing Automation, Steel Microstructure,  \\
             $20 \pm 3$ &          Machine Learning &  Machine Learning, Classification, Support Vector Machines, Decision Tree,  \\
             $20 \pm 1$ &                Microbiome &  Gut Microboime, Fecal Microbiota, Bacteria, Bacterial Community, \\
             $19 \pm 4$ &        Infectious Disease &                                                          Antibiotics, Viruses, Dengue \\
             $19 \pm 2$ &                 Materials &  Molybdenum Sulfide Monolayers, Metal-Organic Frameworks,  \\
             $18 \pm 2$ &           Portable Energy &  Supercapacitor, Batteries, Lithium Sulfur Batteries, Lithium Batteries, s \\
             $17 \pm 3$ &             Miscellaneous &                                                                   Telescope, Fracking \\
             $17 \pm 0$ &  Unmanned Aerial Vehicles &                             Unmanned Aerial Vehicles, Drones, Umanned Aerial Vehicles \\
             $16 \pm 3$ &               Painkillers &                                                             Pain, Opioid, Bupivacaine \\
             $15 \pm 1$ &                   Farming &                                                       Farming, Crop Yield, Fertilizer \\
             $15 \pm 0$ &                    Health &  Smoking, Caregiving, Elderly, Breastfeeding, Geriatric, Tobacco, Dementia,  \\
             $15 \pm 1$ &              Urbanization &               Autonomous Vehicles, Urbanization, Pollution, Wastewater, Air Pollution \\
             $14 \pm 1$ &            Electric Grids &                                           Microgrids, Grid, High Voltage DC Converter \\
             $14 \pm 2$ &        Sports Performance &                                                   Track \& Field, Exercise, Concussion \\
\bottomrule
\end{tabular}

\end{table*}
% ==================

\subsubsection{Areas of Specialty}

% **********************************
% FIGURE:  USA v CHINA LQ
% **********************************
\begin{figure}
\includegraphics[width=\columnwidth]{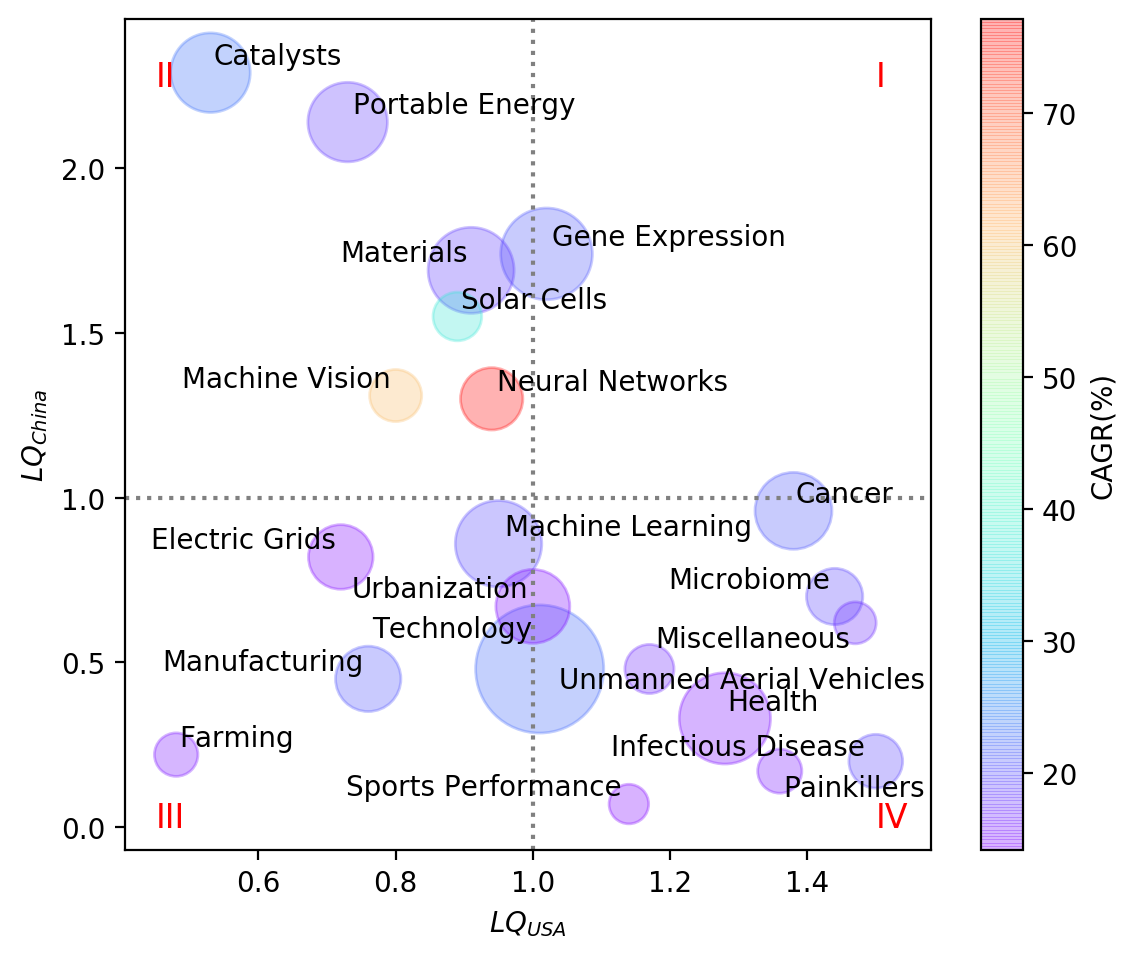}
\caption{A plot of the \textit{Location Quotients (LQ)} of the USA and China for the fastest growing 200 $(2\%)$ topics aggregated into 20 super topics. \textit{Quadrant I,III}: Topics where USA and China are both well represented, or both poorly represented, respectively. \textit{Quadrant II,IV}: Topics where either USA or China are more represented than the other, respectively.}
\label{fig_lq_quad_usa_china}
\end{figure}
% **********************************

While LQ can be calculated over many different category types in the data set, specifically aggregating over \textit{topics} and/or \textit{supertopics} allows LQ to be interpreted as an entity's areas of topic focus, or specialization. An example using this metric is shown in Figure \ref{fig_lq_quad_usa_china}, which plots the LQ for USA and China for the super topics shown in Table \ref{table_supertopics}.  Each quadrant of this chart indicates areas of specialty.  While the USA leads in the most areas, largely in health topics like \textit{cancer, infectious disease}, and \textit{microbiome}, China focuses on fewer topics, with clear leads in \textit{catalysts} and \textit{portable energy}.  China also maintains a slight lead in AI: \textit{neural networks} and \textit{machine learning}.  Both the USA and China have limited relative focus in \textit{farming, manufacturing and electric grids}. The fastest growing areas, \textit{machine vision, deep learning and solar cells}, are within China's quadrant but relatively close to the center $(1,1)$, which indicates these may be areas of competition. A similar plot with similar conclusions can be constructed using authors, author organizations, funding sponsors, etc. to understand the areas of focus of all of these entities relative to each other.

% =================================================
%           DISCUSSION
% =================================================

\subsection{Applications}

\subsubsection{Technology Scouting}

Horizon Scanning for Emerging Technologies is a popular exercise in the public sector\cite{popper_foresight}. Understanding the landscape of emerging technologies and their potential threats drive decisions on how resources are allocated. This methodology nominates emerging technology areas for their consideration, which can then be evaluated by subject matter experts and analysts to assess their disruptive potential. Using LQ in combination with CAGR at the country level can give some indication of rising technological threats, for example what is shown in Figure \ref{fig_lq_quad_usa_china}. Ultimately, this could inform how current resources are allocated and provide an early warning system as to what the threat may possibly be, and from where it will come from.

For government R\&D funding organizations heavily engaged in the strategic investment in various science and technology areas, understanding the landscape of major players, whether they be state actors, similar funding authorities, or research organizations is an important starting point for managing their research investment portfolio. LQ can provide information on the specialization of various entities, which can guide their decision making. For example, they can be valuable when government R\&D organizations are considering new research initiatives, or as a means of rapidly understanding a particular organization's strengths during a competitive selection. Using the LQ on other grant awarding organizations, based on sponsorship mentions in publication acknowledgments, can also give the government R\&D organization insight into what other government agencies are also supporting particular topic areas. This could present the opportunity for these organizations to work cooperatively.

\section{Future Work}

Due to space limitations, a full analysis and exploration of the results combined with analyst interpretations of this horizon scan will be the topic of a future report. The use of a dynamic model of topical areas offers the potential of forecasting the dominance of currently emerging science and technology areas. However, our use of first-order rate kinetics is limited to only a few years since its assumptions break down at longer timescales - thus future work will focus on developing a more comprehensive model.

%% ===========================================================================================
%%     ACKNOWLEDGEMENTS
%% ===========================================================================================

\section{Acknowledgments}
Some of this work was generated in the course of supporting the \textit{Data and Analytics Laboratory at the Office of Naval Research (ONR)} through contract number GS00Q09BGD0019 in their mission to develop better tools for research portfolio analytics. Cleared by ONR for public release: DCN: 43-895221.

%% ===========================================================================================
%%     BIBLIOGRAPHY
%% ===========================================================================================

\bibliographystyle{acm}
\bibliography{ref}

\begin{thebibliography}{10}

\bibitem{for}
The australian and new zealand standard research classification - fields of
  research.

\bibitem{dash}
Dash by plotly.

\bibitem{dimensions}
Digital science. dimensions | the next evolution in linked scholarly
  information.

\bibitem{bokeh}
Interactive data visualization in the browser, from python.

\bibitem{hsdef_oecd}
Overview of methodologies.

\bibitem{policy_roadmap}
{\sc Ahlqvist, T., Valovirta, V., and Loikkanen, T.}
\newblock Innovation policy roadmapping as a systemic instrument for
  forward-looking policy design.
\newblock {\em Science and Public Policy 39}, 2 (2012), 178--190.

\bibitem{hs_review}
{\sc Amanatidou, E., Butter, M., Carabias, V., K{\"o}nn{\"o}l{\"a}, T., Leis,
  M., Saritas, O., Schaper-Rinkel, P., and van Rij, V.}
\newblock On concepts and methods in horizon scanning: Lessons from initiating
  policy dialogues on emerging issues.
\newblock {\em Science and Public Policy 39}, 2 (2012), 208--221.

\bibitem{weak_signals}
{\sc Ansoff, H.~I.}
\newblock Managing strategic surprise by response to weak signals.
\newblock {\em California management review 18}, 2 (1975), 21--33.

\bibitem{bettencourt_emergence}
{\sc Bettencourt, L., Kaiser, D., Kaur, J., Castillo-Chavez, C., and Wojick,
  D.}
\newblock Population modeling of the emergence and development of scientific
  fields.
\newblock {\em Scientometrics 75}, 3 (2008), 495--518.

\bibitem{blei_lda}
{\sc Blei, D.~M., Ng, A.~Y., and Jordan, M.~I.}
\newblock Latent dirichlet allocation.
\newblock {\em Journal of machine Learning research 3}, Jan (2003), 993--1022.

\bibitem{boyack_citation_analysis}
{\sc Boyack, K.~W., and Klavans, R.}
\newblock Co-citation analysis, bibliographic coupling, and direct citation:
  Which citation approach represents the research front most accurately?
\newblock {\em Journal of the American Society for information Science and
  Technology 61}, 12 (2010), 2389--2404.

\bibitem{hs_industry}
{\sc Brown, D.}
\newblock Horizon scanning and the business environment—the implications for
  risk management.
\newblock {\em BT Technology Journal 25}, 1 (2007), 208--214.

\bibitem{matplotlib}
{\sc Caswell, T.~A., Droettboom, M., Lee, A., Hunter, J., Firing, E., Stansby,
  D., Klymak, J., Hoffmann, T., de~Andrade, E.~S., Varoquaux, N., Nielsen,
  J.~H., Root, B., Elson, P., May, R., Dale, D., Lee, J.-J., Seppänen, J.~K.,
  McDougall, D., Straw, A., Hobson, P., Gohlke, C., Yu, T.~S., Ma, E., Vincent,
  A.~F., Silvester, S., Moad, C., Kniazev, N., Ivanov, P., Ernest, E., and
  Katins, J.}
\newblock matplotlib/matplotlib v3.2.0rc3, Feb. 2020.

\bibitem{chen_citespace}
{\sc Chen, C.}
\newblock Citespace ii: Detecting and visualizing emerging trends and transient
  patterns in scientific literature.
\newblock {\em Journal of the American Society for information Science and
  Technology 57}, 3 (2006), 359--377.

\bibitem{cagr}
{\sc Choi, D.~G., Lee, H., and Sung, T.-k.}
\newblock Research profiling for ‘standardization and innovation’.
\newblock {\em Scientometrics 88}, 1 (2011), 259--278.

\bibitem{termite}
{\sc Chuang, J., Manning, C.~D., and Heer, J.}
\newblock Termite: Visualization techniques for assessing textual topic models.
\newblock In {\em Proceedings of the international working conference on
  advanced visual interfaces\/} (2012), pp.~74--77.

\bibitem{mallet}
{\sc Druck, G., mimno, McCallum, A., Badenes-Olmedo, C., Sutton, C., Claire,
  Singh, S., Yao, L., Mendes, S.~P., Wunderlich, M., Körner, M., Soergel, D.,
  Ring, D., mihaiiancu, Huang, M., drevicko, Capdevila, C., Rutherford, T.,
  Mishra, S., Southern, S., Richardet, R., Rockweiler, N., Hussain, M., Harris,
  J.~D., Chen, J., Turri, G., and Schnober, C.}
\newblock Mallet, Nov. 2019.

\bibitem{adjust_text}
{\sc Flyamer, I., Colin, Xue, Z., Li, A., Vazquez, V., Morshed, N., Neste,
  C.~V., scaine1, and mski\_iksm}.
\newblock Phlya/adjusttext: Trying zenodo, Nov. 2018.

\bibitem{glanzel_labeling}
{\sc Gl{\"a}nzel, W., and Thijs, B.}
\newblock Using ‘core documents’ for detecting and labelling new emerging
  topics.
\newblock {\em Scientometrics 91}, 2 (2012), 399--416.

\bibitem{lq}
{\sc Isserman, A.~M.}
\newblock The location quotient approach to estimating regional economic
  impacts.
\newblock {\em Journal of the American Institute of Planners 43}, 1 (1977),
  33--41.

\bibitem{cluto}
{\sc Karypis, G.}
\newblock Cluto-a clustering toolkit.
\newblock Tech. rep., MINNESOTA UNIV MINNEAPOLIS DEPT OF COMPUTER SCIENCE,
  2002.

\bibitem{hs_eu}
{\sc K{\"o}nn{\"o}l{\"a}, T., Salo, A., Cagnin, C., Carabias, V., and
  Vilkkumaa, E.}
\newblock Facing the future: Scanning, synthesizing and sense-making in horizon
  scanning.
\newblock {\em Science and public policy 39}, 2 (2012), 222--231.

\bibitem{error_propagation}
{\sc Ku, H.~H., et~al.}
\newblock Notes on the use of propagation of error formulas.
\newblock {\em Journal of Research of the National Bureau of Standards 70}, 4
  (1966).

\bibitem{umap}
{\sc McInnes, L., Healy, J., Saul, N., and Großberger, L.}
\newblock Umap: Uniform manifold approximation and projection.
\newblock {\em Journal of Open Source Software 3}, 29 (2018), 861.

\bibitem{mimno_phrasing}
{\sc Mimno, D.}
\newblock Using phrases in mallet topic models.

\bibitem{coherence}
{\sc Mimno, D., Wallach, H.~M., Talley, E., Leenders, M., and McCallum, A.}
\newblock Optimizing semantic coherence in topic models.
\newblock In {\em Proceedings of the conference on empirical methods in natural
  language processing\/} (2011), Association for Computational Linguistics,
  pp.~262--272.

\bibitem{lmfit}
{\sc Newville, M., Otten, R., Nelson, A., Ingargiola, A., Stensitzki, T.,
  Allan, D., Fox, A., Carter, F., Michał, Pustakhod, D., Ram, Y., Glenn, Deil,
  C., Stuermer, Beelen, A., Frost, O., Zobrist, N., Pasquevich, G., Hansen, A.
  L.~R., Spillane, T., Caldwell, S., Polloreno, A., andrewhannum, Zimmermann,
  J., Borreguero, J., Fraine, J., deep 42-thought, Maier, B.~F., Gamari, B.,
  and Almarza, A.}
\newblock lmfit/lmfit-py 1.0.0, Dec. 2019.

\bibitem{popper_foresight}
{\sc Popper, R., et~al.}
\newblock Foresight methodology.
\newblock {\em The handbook of technology foresight\/} (2008), 44--88.

\bibitem{pyldavis}
{\sc Sievert, C., and Shirley, K.}
\newblock Ldavis: A method for visualizing and interpreting topics.
\newblock In {\em Proceedings of the workshop on interactive language learning,
  visualization, and interfaces\/} (2014), pp.~63--70.

\bibitem{hierarchie}
{\sc Smith, A., Hawes, T., and Myers, M.}
\newblock Hi{\'e}rarchie : Interactive visualization for hierarchical topic
  models.

\bibitem{coauthor_networks}
{\sc Sun, Y., Barber, R., Gupta, M., Aggarwal, C.~C., and Han, J.}
\newblock Co-author relationship prediction in heterogeneous bibliographic
  networks.
\newblock In {\em 2011 International Conference on Advances in Social Networks
  Analysis and Mining\/} (2011), IEEE, pp.~121--128.

\end{thebibliography}

%% ===========================================================================================
%%     SUPPLEMENTARY INFORMATION
%% ===========================================================================================

\clearpage
\newpage
\appendix
\newpage
%\onecolumn
\section{Supplementary Information}

A summary of the data, tools and methods used in this work appears in Table \ref{table_tool_summary}.

% ==================
%  TABLE: Tools Summary
% ==================
\begin{table*}[ht]
\small
\caption{Summary of data, tools, software packages, addons and libraries employed in the analyses used in this work.}
\label{table_tool_summary}
 \begin{tabular}{rl}
   \toprule
\textbf{Data/Analytic Method}      &    \textbf{Package: Addons}\\
   \midrule
Data Source                        &    \textit{Dimensions}\cite{dimensions}\\
Topic Modeling/LDA                 &    \textit{MALLET\cite{mallet}}\\
Location Quotient (LQ)             &    \textit{Python 3.81}\\
Compound Annual Growth             &    \textit{Python 3.81: LMFIT\cite{lmfit}}\\
Rate (CAGR)\\
Topic Map Layout                   &    \textit{UMAP\cite{umap}}\\
Topic Map Visualization            &    \textit{Plotly Dash\cite{dash}, Bokeh\cite{bokeh}}\\
                                   &    \textit{Python 3.81: matplotlib\cite{matplotlib}}, \\
                                   &    \textit{adjustText\cite{adjust_text}}\\
   \bottomrule
 \end{tabular}
\end{table*}
% ==================

\subsection{Junk Topics}

Examples of junk topics excluded in this study appear in Table \ref{table_junk}. An explanation for the rejection of each topic appears below with the cooresponding topic index:

\begin{itemize}

\item [8203:] Database artifact example: abstracts which included Indonesian words 

\item [4593:] Example of broad, technical topics that do not specifically indicate a particular technology area.

\item [8260] and 7037: Some topics that are focused on geographic regions.

\item [4551:] Example of a non-technical topic that may be relevant to other disciplines but out of scope for this study.

\item [7375:] Example of a topic area that is growing in size, but due to largely non-technology reasons.

\item [7190:] Low coherency topics are often not specific enough to indicate a technical area.

\item [9548:] The largest topic in this study does not contain any specific terms.

\end{itemize}

% ==================
%  TABLE: Junk Topics
% ==================
\begin{table*}[ht]
\small
\caption{Examples of topics excluded from the analysis}
\label{table_junk}

\begin{tabular}{rllrrc}
\toprule
 \textbf{Topic} &     \textbf{Topic Name} &                                \textbf{Top Terms} &  \textbf{Coherence} &  \textbf{Documents} & \textbf{CAGR(\%)} \\
 \textbf{Index}\\
\midrule
                 8203 &     Indonesian Articles &               dan, indonesia, pada, bali, keyword &                -555 &                2153 &              $65 \pm 3$ \\
                 4593 &                 Placebo &     placebo, randomize, trial, efficacy, baseline &                -407 &                1093 &             $48 \pm 27$ \\
                 8260 &                 Ukraine &         ukraine, ukrainian, exp, national, modern &                -742 &                1526 &              $35 \pm 6$ \\
                 4551 &                   Islam &            islamic, muslim, islam, religious, qur &                -531 &                2841 &              $30 \pm 3$ \\
                 7037 &                Learning &            student, learning, class, teach, skill &                -430 &                1846 &             $99 \pm 14$ \\
                 7375 &         Syrian Refugees &       refugee, asylum, syrian, humanitarian, camp &                -423 &                1672 &              $27 \pm 4$ \\
                 7190 &          [not coherent] &          dock, station, fenugreek, nectin, cancel &               -2040 &                 663 &             $36 \pm 19$ \\
                 4764 &  [administrative terms] &  correction, erratum, online, corrigendum, figure &                -673 &                5677 &             $36 \pm 10$ \\
                 9548 &     [nonspecific terms] &        phenomenon, avoid, call, unexpected, wrong &                -762 &               14039 &               $3 \pm 0$ \\
\bottomrule
\end{tabular}

\end{table*}
% ==================

\subsection{Topic Modeling Parameters}

Tuning files used in \textit{MALLET} were developed by several analysts including the following: (1) Multi-word stoplists were developed based on common non-technical phrases found in the abstracts \textit{i.e. copyright John Wiley and Sons}. (2) A typical lemmatization file was modified by analysts to eliminate false matches \textit{i.e. replacement of “ground” with “grind” was removed so that “ground truth” is not replaced with “grind truth.”} (3) Analyst-curated multi-word replacement files were used to destructively aggregate technical phrases such as \textit{“global positioning system”} to \textit{“global\textunderscore positioning\textunderscore system.”} These phrases were discovered through repeated topic modeling iterations of the corpus and examining the topic phrase output. 

To improve the computation speed as well as make the model more robust to noise (such as infrequently found tokens arising from text markup languages or other text artifacts), the vocabulary was further pruned based on the \textit{inverse document frequency (IDF)} to a fixed vocabulary size of $200,000$ tokens. All terms that occurred in more than $5\%$ of the corpus were eliminated. The fixed vocabulary size and upper bound resulted in a lower bound cutoff of $37$ occurrences.

\begin{figure}
\includegraphics[width=\columnwidth]{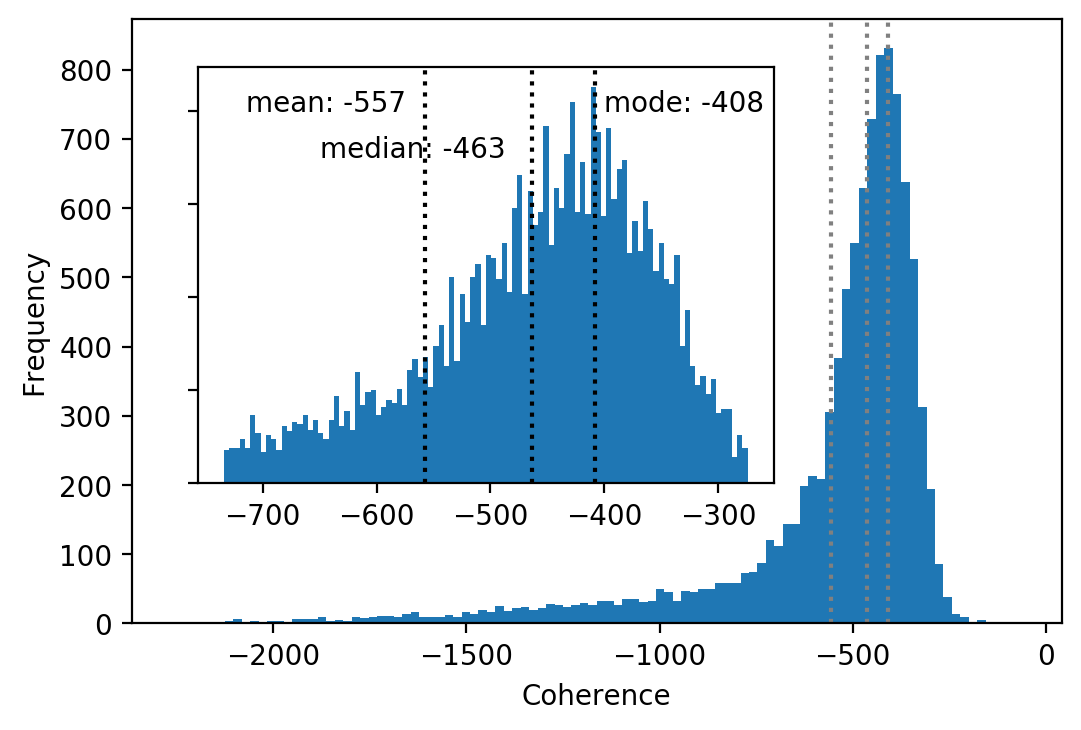}
\caption{A histogram of the \textit{Topic Coherence} for all 10,000 topics modeled by \textit{MALLET LDA}, with the mean, median and mode of the distribution plotted as reference lines. INSET: Expanded view of distribution.}
\label{fig_topic_diagnostics}
\end{figure}

The following parameters were used in MALLET: (1) To allow for different topic sizes, hyper-parameter optimization was performed and carried out every $10$ iterations. (2) Based on examining the \textit{log likelihood per token (LL/token)} generated as part of MALLET's output, $750$ iterations was selected to balance computation time with accuracy. The final optimized \textit{LL/token} was $-7.45623$. (3) Multi-threading was turned on to further improve computation time. Overall, using a Linux-based system with $192$GB RAM and $96$ cores with a computation time of 2 days and 5 hours was achieved. A histogram of the topic coherence for all 10,000 topics appears in Figure \ref{fig_topic_diagnostics}.

% **********************************
% FIGURE:  COHERENCE
% **********************************
% **********************************

\subsection{Evaluation of Fit Quality}

% **********************************
% FIGURE:  EXAMPLE FIT
% **********************************
\begin{figure}
\includegraphics[width=\columnwidth]{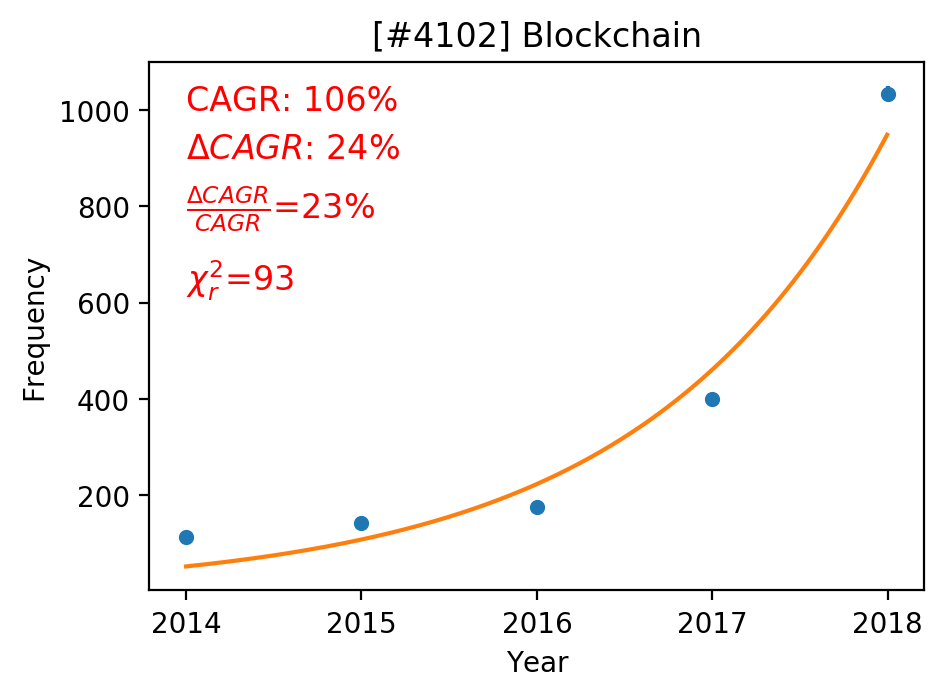}
\caption{Example of a particularly poor fit. Documents (Frequency) versus Year Published for \textit{Topic 4102: Blockchain}, including error bars (which are very similar to the size of the marker). Orange line is the best fit line based on non-linear regression.}
\label{fig_example_fit}
\end{figure}
% **********************************

Figure \ref{fig_example_fit} shows a trace of a particularly poor model fit, \textit{Topic 4102}, which has good coherence and has terms related to \textit{blockchain technology}. Despite the large errors (both in CAGR and the reduced chi-squared, $\chi^2_r$), the model does not appear to do too poorly a job based on visual inspection: it captures the upward trend and overall shape of the data, coming \textit{"quite close"} to all points in the curve. However, because the fitting is many error bars away from the actual points, this is negatively reflected in both the CAGR error and $\chi^2_r$. 

\subsubsection{Document Frequency Error Bars}

Since error rates in document frequencies are difficult to estimate, the following criteria was used to estimate them: (1) they were set proportional to $(\sim \sqrt{N})$, where $N$ is the document count, to reflect the fact that large document counts provide more statistically significant information $\left(\frac{\delta N}{N}=\frac{1}{\sqrt{N}}\right)$ and (2) a scaling factor was determined by iteratively inspecting the $\chi^2_r$ distribution and adjusting to retrieve a $\chi^2_r$ distribution with a mode centered at $1.0$. This procedure avoided having too many $\chi^2_r$s much smaller than 1 (overestimated error bars), or automatically implying the wrong model without further evaluation.

% **********************************
% FIGURE:  CAGR ERR v REDUCED CHI SQUARED
% **********************************
\begin{figure}
\includegraphics[width=\columnwidth]{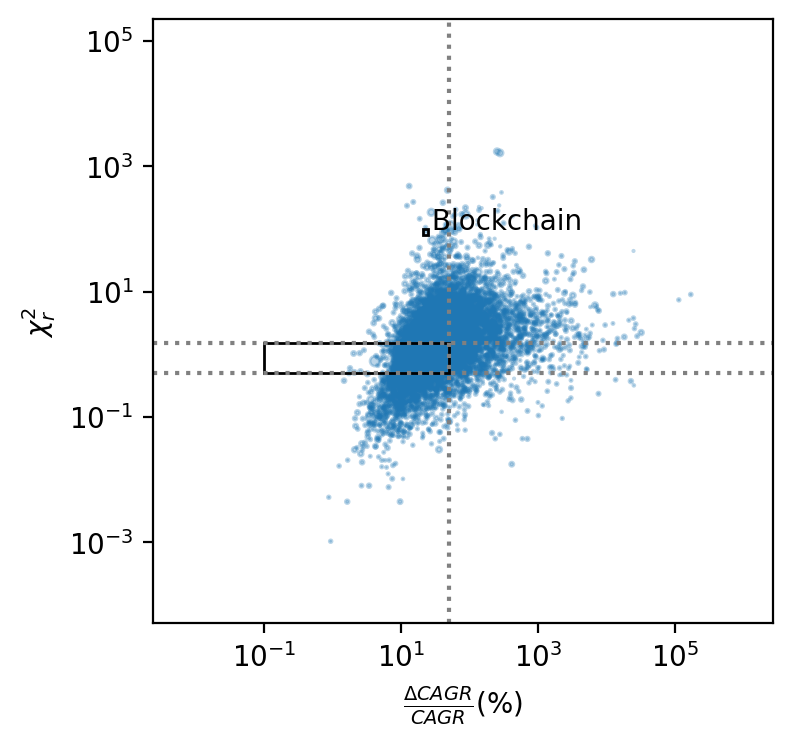}
\caption{Plot of Reduced Chi Squared versus the Percent Error of CAGR. Solid rectangle: a guide to the eye, of good topic model fits (see description in text). For reference, the location of topic 4102: Blockchain is also shown.}
\label{fig_cagr_err_chi}
\end{figure}
% **********************************

\subsubsection{Analysis of Error Rate Distributions}

Figure \ref{fig_cagr_err_chi} is a plot of the CAGR percentage error ($100\times \frac{\Delta CAGR}{CAGR}$) versus ($\chi^2_r$).
The distribution of fits across both parameters is quite broad. Horizontal dotted lines define an arbitrary band of acceptable fits ($0.5<\chi^2_r<1.5$), while the vertical reference line corresponds to CAGR percentage errors less than $50\%$. The topic model fits that satisfy both conditions reside in the "good neighborhood" (small rectangle indicated in figure). 
Points inside the good neighborhood can be viewed as high confidence points (good nonlinear regression and small CAGR percentage error). These topics can have any growth rate but what is emphasized here is the reliability of the results in that good neighborhood. 
Points above the rectangle but to the left of the $50\%$ error (for example the point corresponding to the “Blockchain” topic indicated in the figure) could still be significant: even though their $\chi^2_r$ is large, this could be due to the difficulty in estimating the errors in the time series. Only $1,132$ of the $10,000$ (or $11\%$) topics modeled can be found in the "good neighborhood." 
From a rigorous, and purely statistical standard, the strict interpretation is that the hypothesis that the model explains the majority of the data must be rejected. This can be explained by (1) the small number of data points ($5$ per topic), (2) the simplicity of our exponential growth model (i.e. no constant background, no leveling off), and (3) no good estimates of document frequency errors are, requiring us to apply the above described methodology to estimate them. A more balanced interpretation is that despite these obstacles, it is encouraging that this approach did not completely fail (i.e. capturing a surprising number of good fits ($11\%$) despite the model simplicity), and that even a very poor fit appears to be visually acceptable (Figure \ref{fig_example_fit}). This is a promising foundation for a more detailed growth model that captures the complexity of the scientific enterprise that should be investigated in future work.

%% ===========================================================================================
%%     END MANUSCRIPT
%% ===========================================================================================

\end{document}